\documentclass[11pt,a4paper,onecolumn]{article}
\usepackage[utf8]{inputenc}
\usepackage{amsmath}
\usepackage{amsfonts}
\usepackage{amssymb}
\usepackage{graphicx}
\usepackage[super,sort&compress,comma]{natbib} 
\usepackage[compact]{titlesec}
\usepackage{authblk}
\usepackage[left=2cm,right=2cm,top=2cm,bottom=2cm]{geometry}
\usepackage{pdfpages}

\title{Contribution of \textit{operando} X-ray experiments and modeling to the understanding of the heterogeneous lithiation of graphite electrodes}
\author[1]{Samuel Tardif}
\author[2]{Nicolas Dufour}
\author[2]{Jean-Francois Colin}
\author[2]{G\'erard G\'ebel}
\author[3]{Manfred Burghammer}
\author[3]{Andreas Johannes}
\author[1]{Sandrine Lyonnard}
\author[2]{Marion Chandesris}
\affil[1]{Department of Physics, IRIG, Univ. Grenoble Alpes and CEA, F-38000 Grenoble, France; E-mail: samuel.tardif@cea.fr}
\affil[2]{Department of Electricity and Hydrogen for Transport, Univ. Grenoble Alpes and CEA-LITEN, F-38000 Grenoble, France}
\affil[3]{European Synchrotron Radiation Facility (ESRF), BP 220, Grenoble 38043, France}
\date{}                     
\begin{document}
\maketitle

\section*{Broader Context}
Graphite, alone or blended with silicon, is by far the most used anode material in lithium-ion batteries. To safely enable the fast charge of the high energy density cells designed for the electric vehicle market, one must control the lithium concentration gradients that form across the electrode thickness. Simulations of this phenomenon have been reported in the literature. Experimental validation at the local scale remains however hampered by the required spatial and temporal resolutions. Here, we report on an operando synchrotron x-ray diffraction experiment that fulfills these requirements to measure locally and quantitatively the Li$_{x}$C$_{6}$ phases formed during cycling. The lithium concentration evolution across the electrode thickness follows a very particular pattern with alternating sequences of homogeneous and heterogeneous lithium distributions. The comparison with simulation results comforts modeling predictions at low Li content, but calls for revisiting the balance between the physical driving forces at higher stoichiometries. Experimental observations and theoretical predictions are reconciled by considering a reduction of the insertion kinetics during the LiC$_{6}$/LiC$_{12}$ transition. This finding may have important consequences as negative potentials, which favors the formation of lithium deposits, would be reached for much lower electrode thickness and C-rate than what is currently predicted with classical models.

\section*{Abstract}
Distributions of potential and lithium content inside lithium ion batteries highly affects their performance and durability. 
An increased heterogeneity of the lithium distribution is expected in thick electrodes with high energy densities or cycling at high currents. 
To optimize electrodes and cells designs, it is crucial to probe lithium concentration gradients across the depth of the electrode, but also to predict their occurrence and magnitude as a function of materials properties. 
Here, we follow the lithium distribution across a $80~\mu m$ thick porous graphite electrode using a $1~\mu m$  focused synchrotron X-ray beam. 
The sequential formation of the individual Li$_x$C$_6$ phases during lithium de-insertion is extracted from X-ray diffraction patterns, allowing the quantification of lithium concentration across the electrode thickness. 
Analyzing the evolution of heterogeneities as a function of time, we recover the striking features we predicted with a porous electrode model, including the succession of homogeneous and heterogeneous distributions of lithium. 
However, a clear difference is obtained at high stoichiometry, with a much more homogeneous distribution than initially predicted. 
Revisiting the interplay between transport and kinetic transfers limitations in the porous electrode model, we suggest that the kinetics of lithium (de)-insertion is highly reduced during the LiC$_6$/LiC$_{12}$ phase transition.

\section{\label{sec:intro}Introduction}
Li-ion batteries (LiB) are now ubiquitous in consumer electronics and there is a strong demand from the automotive sector for devices charging faster (\textit{e.g.} using larger currents) and with a larger capacity (\textit{e.g.} using thicker electrodes). \cite{etacheri_challenges_2011, ding_automotive_2019}
However, large current densities and thick electrodes are expected to increase Li concentration gradients across the electrode.\cite{Harris2013, Danner2016, dufour_lithiation_2018}
Such an heterogeneous intercalation of Li can be highly detrimental to the cells performance and durability. Indeed, the distribution of corresponding overpotentials leads to electrode polarization and may results in either Li plating during charge or drastically reduced capacity. 
As a consequence, some volumes of the electrode may reach higher states of charge and thus age faster, while others may not be used to their full capacity.
Additionally, gradients in Li concentration might induce strain and stress gradients that can result in electrode pulverization and electrical disconnection.\cite{Qi2010}

Predicting the amplitude of such Li concentration gradients inside electrodes is key for cell design optimization but remains challenging.
In fact, these gradients result from the complex interaction between various physical phenomena, \textit{i.e.} lithium (de)-intercalation in the active material, or lithium transport in both the active materials and the electrolyte filling the pores of the electrode.
In particular, the latter is strongly impacted by the thickness, porosity and tortuosity of the electrode.\cite{Gallagher2016}
Theoretical studies can help quantifying the concentration gradients, based on the porous media approach of Newman-type models\cite{Newman1975, Doyle1993} or direct simulations of lithium transport and intercalation in detailed 3D microstructure\cite{Danner2016, Hein2016}. 
However, the reliability of these models strongly depends on the consolidated knowledge of the materials transport and intercalation properties. 
Given the experimental difficulties to measure all parameters individually,\cite{Ecker2015a} simplifications and fittings are often performed. 
The validity of the model is usually checked \textit{a posteriori} by the comparison of the simulated cell voltages with electro-chemical measurements only at the cell scale, since there is a dearth of validation data at the particles and electrode scale.

From the electrode perspective, graphite is the most widely used active material for LiB. However the simulation of lithium (de)-intercalation in graphite active materials remains complex as it happens through a succession of different Li ordering between the graphene planes, leading to several phase transitions, \cite{dahn_phase_1991, billaud_revisited_1996} whose dynamics are still insufficiently understood.\cite{gavilan-arriazu_kinetic_2020}
While most models use a solid solution description and Fick's law to simulate the lithium diffusion inside the graphite, more recent works incorporate the dynamics of phase transitions using multi-layer phase field theory to capture the evolution of the phase fronts.\cite{Smith2017, Chandesris2019} 
These new approaches are still under development at the particles scale and are difficult to transpose to the whole electrode level. 
Nevertheless, some recent works have combined phase field description in active materials with a porous description of the electrode.\cite{Smith2017, Thomas-Alyea2017}
Furthermore, Thomas-Alyea \textit{et al.} have shown that both solid solution and phase field approaches give very similar results for the particle average Li concentration and cell potential in out-of-equilibrium conditions,\cite{Thomas-Alyea2017} thus confirming the relevance of a Newman-type model with a solid-solution description to capture the evolution of these quantities.

Using a Newman-type model and a solid-solution approach, we predicted that lithiation in graphite electrodes occurs with a succession of homogeneous and heterogeneous lithium distributions over the thickness of the electrode, with the maximum heterogeneities happening during the plateaus of the graphite equilibrium voltage curve.\cite{dufour_lithiation_2018}
The corresponding simulated electrochemical results (potential, Li content) were validated at the electrode scale, \textit{i.e.} averaged over the whole electrode, by comparison with experimental cell electrochemical measurements. 
However, it is much harder to confirm the lithium distribution predicted by the simulations at a lower scale, \textit{e.g.} across the thickness of the electrode, as one needs both a local, micron-sized probe and an electrochemical cell that allows the measurements to be performed during battery cycling.

To probe heterogeneous graphite lithiation, several \textit{in situ} and \textit{operando} experiments have been performed using optical microscopy and taking advantage of the different colors of the lithiated graphite phases.\cite{Harris2010, Novak2010, Guo2016} 
This method yields interesting qualitative information about phase front evolution inside graphite electrodes, but the method hardly separates contributions from mixed phases. 
Very recently, more quantitative results have been obtained in an \textit{operando} X-ray diffraction experiment with a depth resolution of 20$\mu$m in the graphite electrode.\cite{yao_quantifying_2019}
Focusing on the quantification of lithium gradients that develop in a porous graphite electrode, Yao \textit{et al.} \cite{yao_quantifying_2019} reported a complex dynamic during the partial charging or discharging of their 114~$\mu$m thick graphite electrode. 
The observed dynamic seemed to include the succession  of homogeneous/heterogeneous Li distributions, although the spatial resolution was not sufficient for a detailed quantification.

Here we report on \textit{operando} microfocused synchrotron X-ray diffraction measurements to quantify the succession of Li$_x$C$_6$ phases formed during a complete (de)lithiation across the thickness of a graphite electrode at the micron scale.
We describe the obtained experimental results, focusing our analysis on the quantification of the heterogeneities that develop through the thickness of the electrode during (de)lithiation. 
The experimental results are confronted to numerical predictions obtained with a porous electrode model and we further discuss how these experimental results imply revisiting some assumptions of the model.
In particular, we propose a new shape for the exchange current density, beyond the classical Butler-Volmer model.

\section{\label{sec:exp}Modeling and experiment}
\subsection{Electrode model and heterogeneity quantification}
We use a graphite electrode model based on the porous electrode theory, originally introduced by Newman and Tiedemann.\cite{Newman1975}
The resolved equations are classical and details about the model calibration and validation can be found in Dufour~\textit{et al.}\cite{dufour_lithiation_2018}
To quantify the heterogeneity of lithium concentration upon cycling, we introduce the Normalized Absolute Averaged Deviation (NAAD):\cite{gu_mathematical_1983}
\begin{equation}
    NAAD = \frac{1}{L}\int_{z=0}^{z=L}\frac{|x_{Li}(z)-\langle x_{Li}\rangle_z|}{\langle x_{Li}\rangle_z}dz
\end{equation}
where $L$ is the electrode thickness, $x_{Li}(z)$ is the lithium concentration at position $z$, and $\langle x_{Li}\rangle_z$ is the mean lithium concentration over the electrode thickness. 
The NAAD is a direct quantification of the lithium homogeneity within the electrode: a small (large) NAAD indicates an homogeneous (heterogeneous) electrode lithiation. 

\begin{figure}
    \centering
	\includegraphics[scale=1]{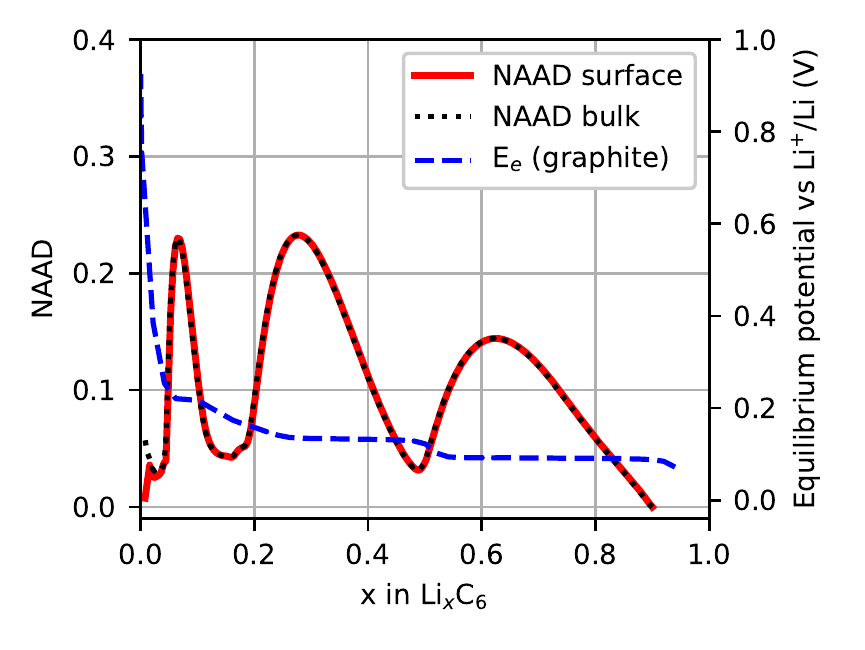}
	\caption{\label{fig:NAAD_volume_surface} Calculated NAAD of the graphite particle Li content, at the surface (thick red line) and in the bulk (black dotted line) as function of $x$ in Li$_X$C$_6$ during a delithiation at C/5, and corresponding equilibrium potential of graphite (E$_{e}$, blue dashed line). (colors online)}
\end{figure}

In numerical models, the NAAD can be computed either for the lithium concentration at the surface of the active particles or for the mean lithium concentration inside the particles, as both quantities can be accessed. 
In Dufour \textit{et al,} we computed the NAAD of the lithium at the surface of the graphite particles ($x_{Li}^s$) while focusing on the mechanisms at the origin of the different resistances inside the electrode.\cite{dufour_lithiation_2018}
We predicted the occurrence of successive heterogeneous and homogeneous distribution during lithiation of the graphite electrode. 
Using the same model, we show here that similar NAAD patterns are obtained for lithium concentration at the surface of the particles and inside the particles (figure~\ref{fig:NAAD_volume_surface}), as long as the diffusion inside the graphite particles is not the limiting factor. 
More precisely, the predicted NAAD pattern has a particular wave shape with three maxima located around $x$ equals 0.06, 0.27 and 0.61, and three minima, located around 0.03, 0.15 and 0.49. 
The $x$ position of these extrema is strongly correlated with the graphite equilibrium potential, also displayed in figure~\ref{fig:NAAD_volume_surface}, which presents three plateaus and three transition regions. 
For states of charge corresponding to a plateau region, there is no modification of the equilibrium potential upon delithiation, thus no potential modification, while transporting lithium from the back of the electrode is a source of overpotential. 
This promotes lithiation disparities along thickness, and thus a maximum NAAD. 
On the contrary, on a transition region between two plateaus, transport of lithium from the back of the electrode can induce less potential differences than the transition from one plateau to the next one, boosting the return to a homogeneous distribution of lithium through the electrode thickness and thus a minimum NAAD.       

\subsection{Cell design and materials}
\begin{figure}
    \centering
    \includegraphics[scale=0.45]{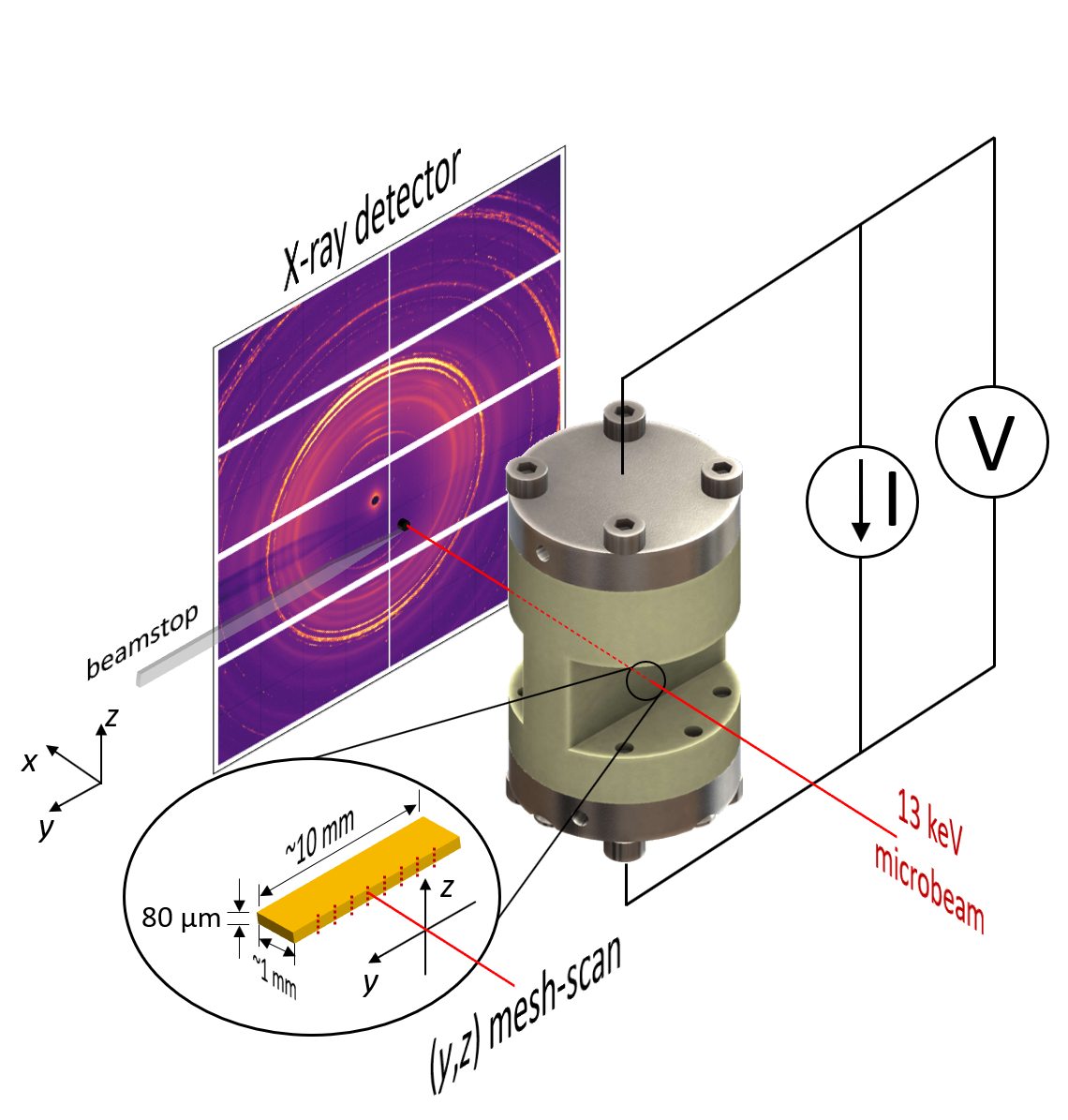}
	\caption{Experimental principle showing the direction of the X-ray beam, the \textit{operando} electrochemical cell with entry and exit windows and a zoom on the electrode, the diffraction pattern recorded behind the cell, as well as the $y$ and $z$ directions along which the cell is moved to scan various locations along the thickness of the graphite electrode. \label{fig:cell}}
\end{figure}

An \textit{operando} electrochemical cell has been designed to perform the local microdiffraction measurements needed to validate the lithium heterogeneities predicted by the model.  
The cell body comprises 200 $\mu$m-thick X-ray transparent polyetheretherketone (PEEK) entry and exit windows, as shown in figure \ref{fig:cell}.
The electrochemical cell was assembled in a glovebox and consisted of the graphite electrode, which is roughly 80~$\mu$m-thick, two microporous Celgard 2500 separators of 25~$\mu$m thickness each, and a 250~$\mu$m-thick metal lithium foil. 
The graphite electrode, which was roughly 1~cm-long in the direction perpendicular to the beam contains 96\%wt of SLP30 graphite powder from TIMCAL, 2.5\%wt conductive carbon particles and 1.5\%wt of polymer binder coated on a 15~$\mu$m-thick copper foil current collector. 
To reduce as much as possible the attenuation of the beam and to limit the parallax effects, the graphite electrode was only 1 mm large in the direction of the beam. 
The areal capacity of the electrode was around 4.5~mAh.cm$^{-2}$. 
A thick lithium foil was used for the counter electrode since its transparency to the beam prevents the diffracted beam from being masked by the upper stainless steel cap while measuring the graphite electrode region close to the separator. 
The cell was filled with 1M LiPF$_6$ in 1:1:1 volume proportion of ethylene carbonate (EC), ethyl methyl carbonate (EMC) and dimethyl carbonate (DMC).

\subsection{Cycling conditions}
The cycling tests have been conducted at room temperature using a VSP potentiostat from Biologic. 
The graphite electrode was cycled once and relithiated prior to the diffraction experiment, the detailed procedure is reported in the Supplementary Information.
It was then delithiated during the diffraction experiment at a constant current of 0.1~mA, corresponding roughly to C/5,  between 0~V and  1.5~V. 

\subsection{Microdiffraction experiment}
The microdiffraction data were collected on the ID13 beamline at the European Synchrotron Radiation Facility (Grenoble, France).
The X-ray beam energy was set to 13~keV  ($\lambda = 0.9537 \text{\AA}$) to reduce the attenuation of the different cell components while maintaining the focusing ability. 
A two-dimensional Dectris EIGER 4M pixel detector was placed 175~mm behind the electrochemical cell.
The exact position of the detector was calibrated prior to the experiment using Al$_{2}$O$_{3}$ powder.
The cell was mounted on a stepper motor to control its exact position, to allow a very fast displacement, and to scan it in both $y$ and $z$ directions (figure~\ref{fig:cell}).  
The microfocus probe size at the sample location was 1~$\mu$m$\times$1~$\mu$m and the cell was repeatedly scanned over a mesh of 8 points along $y$ (width) and 51 points along $z$ (thickness). 
These 8 $\times$ 51 measures will be also referred as one scan, hereafter.
The $z$ step size was 3~$\mu$m and the $y$ step size was 500~$\mu$m.
We used such a large $y$ step and shifted the $y$ points between each mesh by 5~$\mu$m (and circling back every 48 meshes) in order to avoid repeated measurements over the exact same region, \textit{i.e.} to distribute the dose over the electrode. 
An exposure time of 10~ms was sufficient for the acquisition of the diffraction powder pattern at each $(y,z)$ location.
The total time for a complete mesh was about 110~s, \textit{i.e.} including motors movements and deadtime.

\section{\label{sec:level1}Results}
\subsection{Electrode-averaged analysis}
\begin{figure}
    \centering
    \includegraphics[scale=1.0]{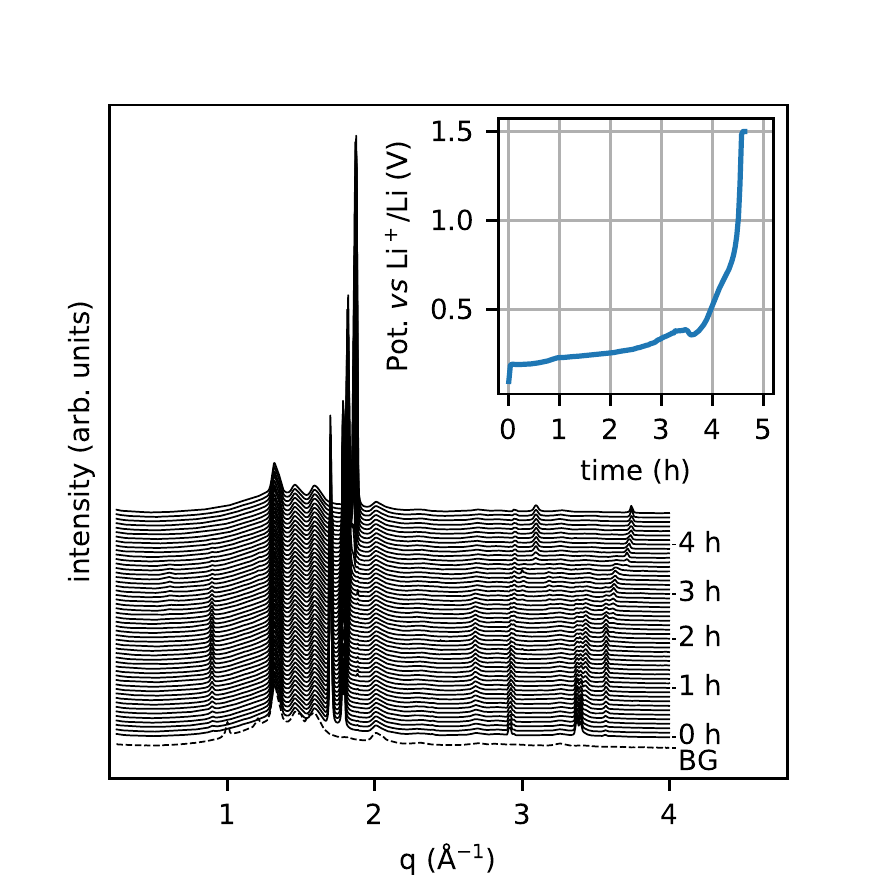}
    \caption{\label{fig:Ech1_deLi1_waterfall}Powder diffraction patterns averaged over the electrode area, as a function of time. They correspond to the average composition of the whole electrode over the scan duration. The patterns have been shifted vertically and only one every three is plotted for clarity. The dashed curve is the background (BG) measured above the electrode, through the separator. (inset) Potential of the half-cell during the delithiation experiment, as a function of time.}
\end{figure}

An overview of the evolution of the state of the whole electrode during the delithiation can be obtained by averaging the $8 \times 51$ measurements of each scan, as shown in figure \ref{fig:Ech1_deLi1_waterfall}.
The high intensity peaks between $1.70$ and $1.87~\text{\AA}^{-1}$ correspond to reflections on successive graphene planes, \textit{i.e.} d-spacing between $3.70~\text{\AA}$ (LiC$_{6}$ $001$) and $3.36~\text{\AA}$ (graphite $002$).
Second higher order reflections (\textit{e.g.} LiC$_{6}$ $002$, graphite $004$) can also be seen at twice the q-values, between $3.40$ and $3.74~\text{\AA}^{-1}$.
Additional peaks with in-plane components ($h$ or $k\ne0$) are visible around $3~\text{\AA}^{-1}$ (\textit{e.g.} LiC$_{6}$ $2\bar{1}0$ at $2.91~\text{\AA}^{-1}$, graphite $110$ at $3.09~\text{\AA}^{-1}$).
The intense peaks as well as the diffuse one between $1.0$ and $1.7 \text{\AA}^{-1}$ stem from the PEEK of the cell windows, as indicated by a measurement above the electrode, through the separator (dashed line in figure \ref{fig:Ech1_deLi1_waterfall}).
Note that if present, Li metal would appear at $2.53~\text{\AA}^{-1}$ $(110)$ or $3.58~\text{\AA}^{-1}$ $(002)$.
The inset in figure \ref{fig:Ech1_deLi1_waterfall} shows the simultaneous evolution of the potential as a function of the elapsed time. %scan number.
As expected for a delithiation reaction, the potential rises from a low voltage value to the higher cut-off potential (1.5~V). 

\begin{figure}
    \centering
    \includegraphics[scale=0.5]{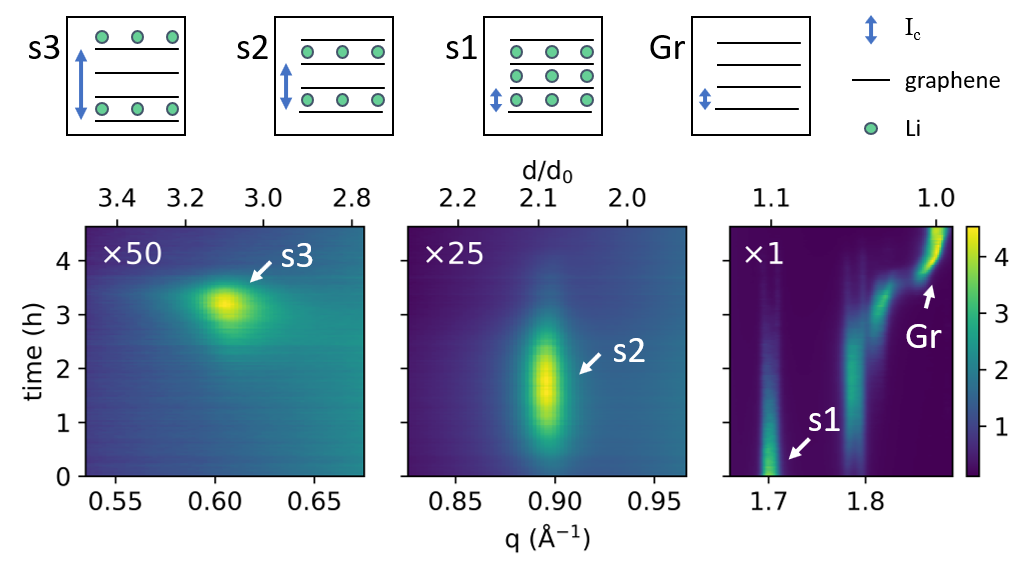}
    \caption{\label{fig:Ech1_deLi1_Imap_details}Intensity map of the diffraction around selected q-values indicative of the stage 3 (left, s3), stage 2 (center, s2), and stage 1 and stage dilute + graphite (right, s1 and Gr respectively). The top scale indicates the corresponding relative d-spacing to that of graphite ($\mathrm{d}_{0} = 3.355~\text{\AA}$). The intensity scale factor is indicated in the top left. The sketches above illustrate the corresponding reflections in the intensity maps}
\end{figure}

The different lithiation stages $n$ can be readily identified by the examination of reflections on planes separated by about $n$ times the graphite interlayer distance (\textit{i.e.} the repetition length $\mathrm{I}_{\mathrm{c}}$ along the ${\mathrm{c}}$ axis), independently of their composition or detailed crystalline structure.
Figure \ref{fig:Ech1_deLi1_Imap_details} shows a map of the intensity around selected $q$ values corresponding to 3, 2 and 1 times the interlayer d-spacing in empty graphite ($\mathrm{d}_{0}=3.355~\text{\AA}$). 
Note that no peak could be resolved above the background level for $n\geq4$.
At the beginning of the measurement ($t = 0$~h), when the graphite electrode is almost fully lithiated, the intensity is maximum around $q = 1.7~\text{\AA}^{-1}$, \textit{i.e.} an interlayer distance of about $1.1\times\mathrm{d}_{0}$.
This indicates a stage 1 lithiation, with the graphite dilated by about 10$\%$ along the c axis.
During the delithiation, this peak fades away and a new peak appears at a d-spacing of $2.1\times\mathrm{d}_{0}$ ($t \approx 0.6$~h to $t \approx 2.5$~h), corresponding to the stage 2.  
Similarly, when this latter peak decreases, another peak appears at a d-spacing of about $3.1\times\mathrm{d}_{0}$ ($t \approx 2.5$~h to $t \approx 3.5$~h), indicating a transition to stage 3. 
Finally the stage 3 disappears, and a new peak appears at a d-spacing slightly larger than $\mathrm{d}_{0}$ and shifts towards $\mathrm{d}_{0}$ by the end of the delithiation ($t = 4.6$~h), marking the transition to the stage 1d (dilute) and then pure graphite. 
As seen on the right panel of figure \ref{fig:Ech1_deLi1_Imap_details}, reflections on successive graphene planes are also seen for stage 2 and 3, at $q =1.79~\text{\AA}^{-1}$ and $1.82~\text{\AA}^{-1}$, corresponding to $\mathrm{d}=\mathrm{I}_{\mathrm{c}}/2$, \textit{i.e.} $\mathrm{d}/\mathrm{d}_{0}=1.05$, and $\mathrm{d}=\mathrm{I}_{\mathrm{c}}/3$, \textit{i.e.} $\mathrm{d}/\mathrm{d}_{0}=1.03$, respectively.

\begin{figure}
    \centering
    \includegraphics[scale=1.0]{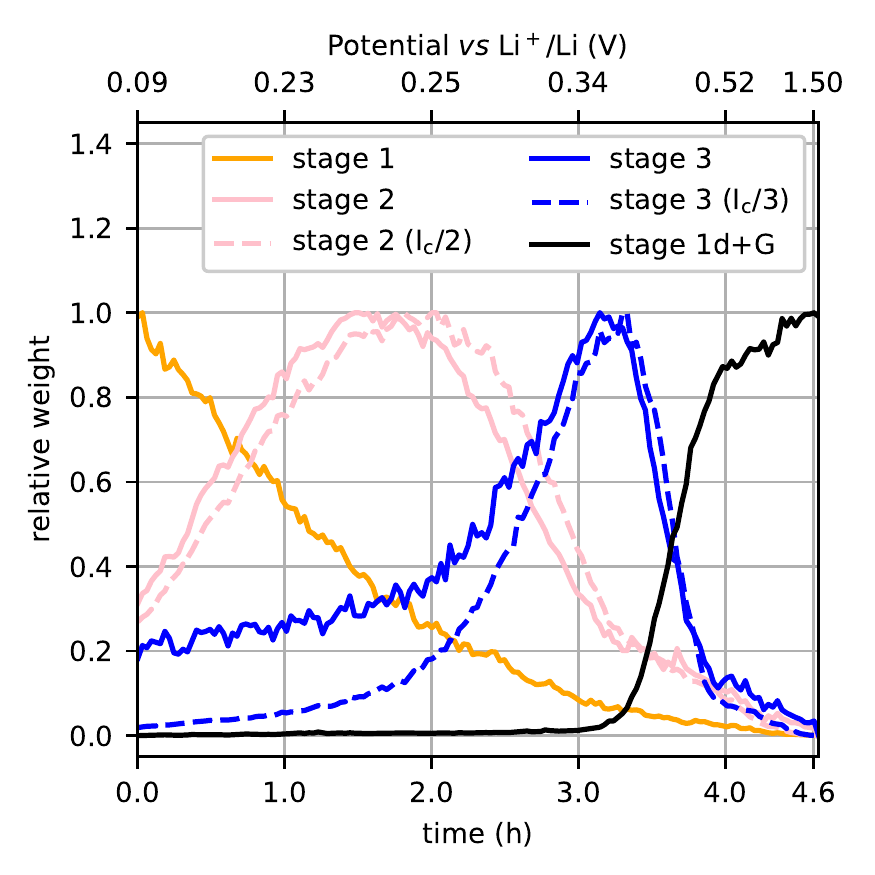}
    \caption{\label{fig:Ech1_deLi1_stage_succession}Relative weight of the integrated intensity around the reflections indicated in the text, corresponding to stage 1, 2, 3 and 1d + graphite (1d+g). The upper x axis indicate the corresponding potential of the cell measured by the galvanostat.}
\end{figure}

Once the different stages have been identified, the relative contribution of each stage to the overall electrode state has been estimated by comparing the relative variations of the integrated intensity around the corresponding $q$ values (stage 1 from $1.680$ to $1.730~\text{\AA}^{-1}$, stage 2 from $0.860$ to $0.930~\text{\AA}^{-1}$, stage 3 from $0.570$ to $0.640~\text{\AA}^{-1}$, stage dilute and graphite from $1.841$ to $1.910~\text{\AA}^{-1}$). 
The corresponding integrated intensities are shown in figure \ref{fig:Ech1_deLi1_stage_succession}, where the minimum and maximum values for each curve have been arbitrarily set to 0 and 1, respectively.
A clear succession of stages of increasing order can be observed during the delithiation, with each stage reaching successively a peak in maximum proportion.
The stage 2 and 3 can also be estimated from the integrated intensity around reflections on successive graphene planes (\textit{i.e.} from $1.740$ to $1.801~\text{\AA}^{-1}$ for $\mathrm{d}=\mathrm{I}_{\mathrm{c}}/2$, and from $1.801$ to $1.841~\text{\AA}^{-1}$ for $\mathrm{d}=\mathrm{I}_{\mathrm{c}}/3$, for stage 2 and 3, respectively). 
The relative maxima of stage 2 and 3 composition calculated in this fashion are slightly shifted towards later times: while the intensity is much larger for those reflections, they are very close together and it can be difficult to separate successive contributions.
The initial composition is mostly stage 1, as expected from the preliminary lithiation of the electrode.
However, the stage 2 proportion is not at its minimum, suggesting that the electrode was not fully lithiated. 
Initial stage 3 content is around zero according to the integrated intensity around $\mathrm{d}=\mathrm{I}_{\mathrm{c}}/3$. 
The non-vanishing initial stage 3 content measured from the fundamental reflection ($\mathrm{q}=0.6~\text{\AA}^{-1}$) stems from the poor signal-to-background ratio for this weak reflection and thus a high sensitivity to small background variations.
In the following we base our analysis on the most intense reflections (q between $1.7$ and $1.9~\text{\AA}^{-1}$). 
In conclusion of the electrode-averaged analysis, the electrochemical performances of the \textit{operando} cell were satisfactory and we observed the expected structural evolution, in agreement with the literature. \cite{guerard_intercalation_1975,trucano_structure_1975,dahn_phase_1991,ohzuku_formation_1993,billaud_revisited_1996,wang_visualizing_2012,vadlamani_-situ_2014,boulet-roblin_operando_2016,taminato_real-time_2016,missyul_xrd_2017,canas_operando_2017, mathiesen_understanding_2019, matsunaga_comprehensive_2019, berhaut_multiscale_2019,berhaut_prelithiation_2020,didier_phase_2020}

\subsection{Depth-resolved analysis}
\begin{figure}
    \centering
    \includegraphics[scale=1.0]{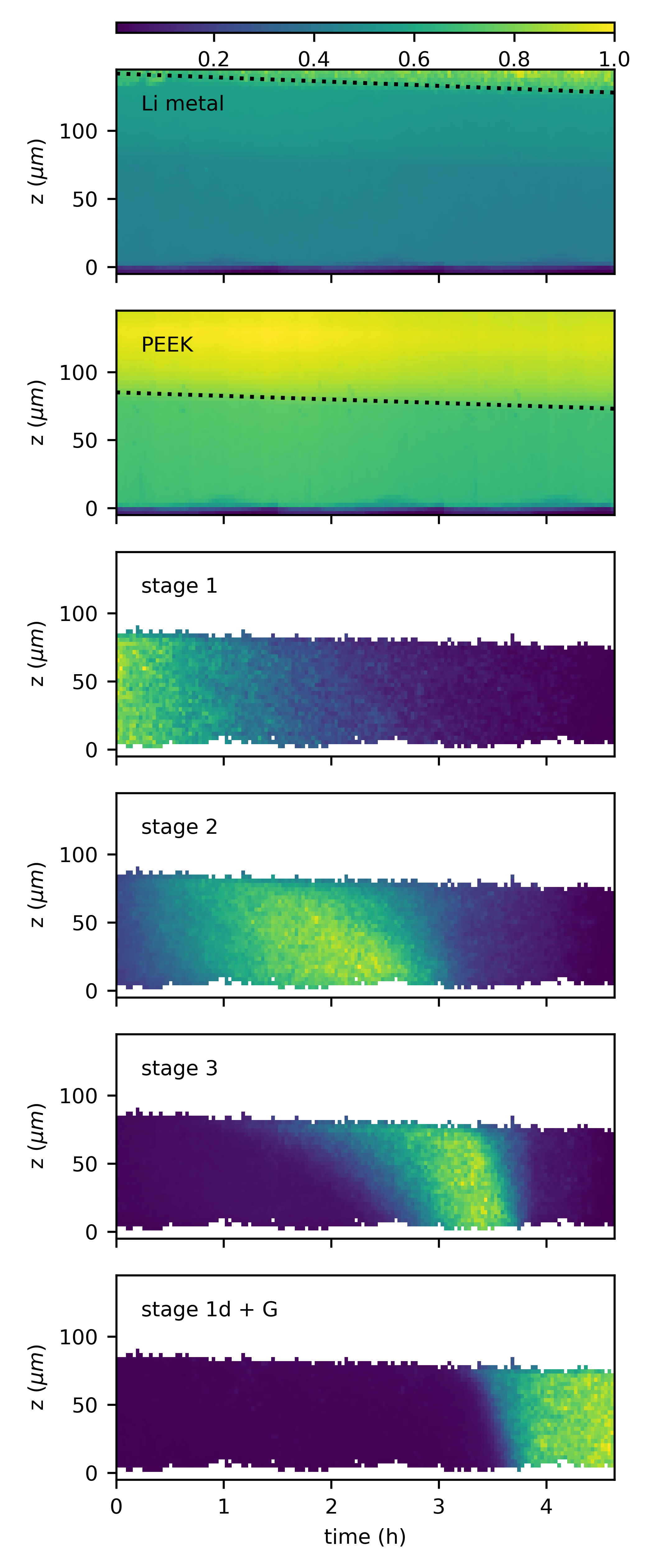}
    \caption{\label{fig:Ech1_deLi1_phases_maps}Color maps of the relative intensity integrated over different $q$ ranges, corresponding to different lithiation stages or phases: Li metal from $2.500$ to $2.560~\text{\AA}^{-1}$, PEEK (cell body) from $1.270$ to $1.630~\text{\AA}^{-1}$, stage 1 from $1.680$ to $1.730~\text{\AA}^{-1}$, stage 2 from $1.740$ to $1.801~\text{\AA}^{-1}$, stage 3 from $1.801$ to $1.841~\text{\AA}^{-1}$, and stage dilute and graphite from $1.841$ to $1.910~\text{\AA}^{-1}$.}
\end{figure}

Now that we have shown the overall evolution of the electrode, we turn to a depth-resolved description.
The depth distribution of each phase can be obtained similarly as before by integrating the intensity over $q$ values around each peak for each $z$ position of each scan, only keeping the averaging along $y$. 
The results are shown in figure \ref{fig:Ech1_deLi1_phases_maps}, using the same normalization between 0 and 1 as in figure \ref{fig:Ech1_deLi1_stage_succession}. 
The Li metal counter electrode can be observed in the top part of the intensity map, and we can resolve the deposition of Li metal during the delithiation as the thickness of the electrode linearly increases over time.
No traces of Li metal could be observed elsewhere, either in the graphite electrode or in the separators.
In the intensity map from the PEEK windows of the cell, the shadow of the more absorbing graphite electrode and Cu current collector are clearly visible.
The distance between the surface of the graphite electrode ($z \approx 80~\mu m$) and the surface of the Li metal ($z \approx 130~\mu m$) corresponds to the thickness of the two $25~\mu m$-thick Celgard separators.
The interface of the graphite electrode with the separator constantly shifts towards lower $z$ values while the interface with the Cu current collector stays constant.
This indicates that the thickness of the graphite electrode decreases over time, while Li is extracted and deposited on the counter electrode, as expected.
The shape of the electrode shadow over time was therefore used as a mask for the intensity map of stages 1, 2, 3, and dilute + graphite. 
In the initial state, the electrode is an uniform stage 1 state, though the stage 2 intensity is not at its minimum, as discussed above.
As the delithiation begins, the stage 1 content decreases while the stage 2 content increases in a seemingly homogenous fashion across the thickness of the electrode, up to about $t = 2$ h. 
Then, the stage 2 content starts decreasing and the stage 3 content starts increasing from the surface.
The transition takes some time to reach the current collector at the bottom of the electrode, as can be seen from the slanted shape of the color map between roughly $t = 2$ h and $t = 3$ h.
Similarly, the stage dilute + graphite starts to appear on the surface while the stage 3 content decreases.
Stage dilute + graphite progresses towards the current collector over time, between roughly $t = 3.5$ h and $t = 4$ h.
The propagation of the transition from stage 3 to stage dilute + graphite from the electrode surface to the current collector is thus faster than the previous one, as seen from the more vertical slope.
At the end of the delithiation, the electrode is in a homogeneous graphite state.

It is clear from these results that the graphite electrode undergoes different distributions of compositions across its thickness during the delithiation.
The distribution of composition can be homogeneous (\textit{e.g.} stage 1, stage 1 + stage 2, or stage dilute + graphite) or heterogeneous (\textit{e.g.}transition stage 2 to stage 3, or transition stage 3 to stage dilute + graphite).
This type of successive homogeneous/heterogeneous composition distribution across a graphite electrode can also be observed in the results reported by Yao \textit{et al.}.\cite{yao_quantifying_2019}

\begin{figure}
    \centering
    \includegraphics[scale=1.0]{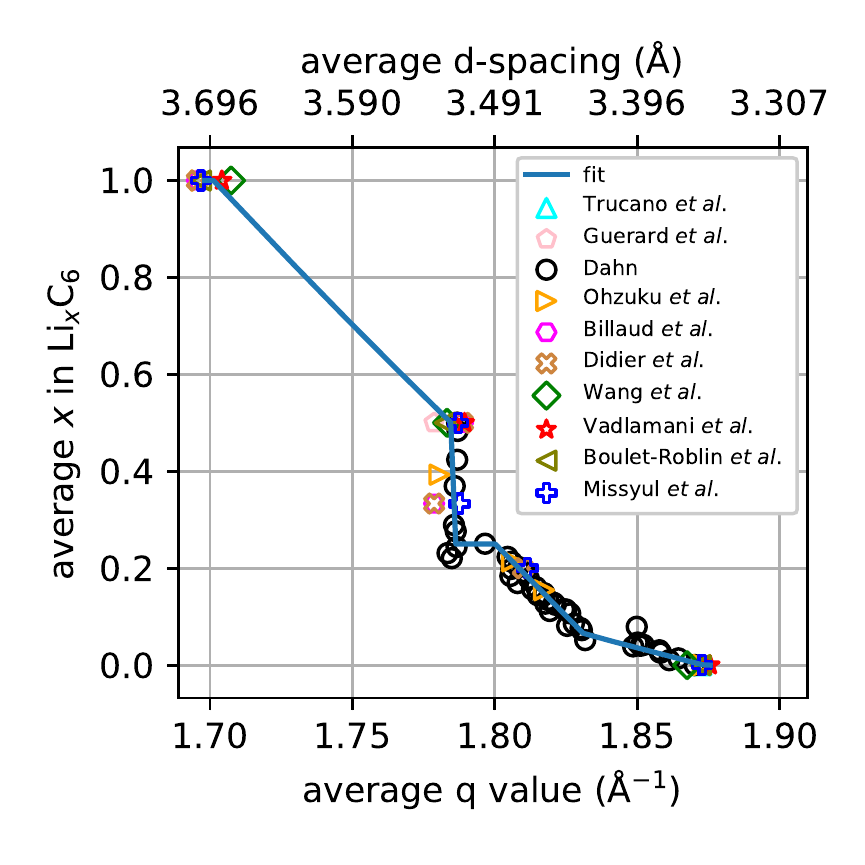}
    \caption{\label{fig:Ech1_deLi1_x_function_of_d}Relationship between the average distance between the graphene sheets (and corresponding $q$ values) and the average electrode Li content $x$ in Li$_{x}$C$_{6}$. 
Symbols are taken from the literature, from reports by Trucano \textit{et al.}\cite{trucano_structure_1975}, Guerard \textit{et al.}\cite{guerard_intercalation_1975}, Dahn \cite{dahn_phase_1991}, Ohzuku \textit{et al.}\cite{ohzuku_formation_1993}, Billaud \textit{et al.}\cite{billaud_revisited_1996}, Wang \textit{et al.}\cite{wang_visualizing_2012}, Vadlamani \textit{et al.}\cite{vadlamani_-situ_2014}, Boulet-Roblin \textit{et al.}\cite{boulet-roblin_operando_2016}, and Missyul \textit{et al.}\cite{missyul_xrd_2017}.
The solid line is least-square fit with a piecewise linear function (see text).}
\end{figure}

In order to go further and describe quantitatively the homogeneous or heterogeneous distribution of the electrode composition, one would require a quantifiable parameter.
The Li content, $x$ in Li$_{x}$C$_{6}$, is of course the most direct metric to quantify the phase composition.
It is however not so easily accessible, and the relation between staging and composition is not straightforward. 
While the stoichiometries of stage 1 and graphite are trivial, respectively LiC$_{6}$ ($x=1$) and C ($x=0$), it is not the case for higher order stages. 
For example, stage 2 is usually further distinguished between stage 2 with an ordered, full Li layer between every second graphene sheet, and stage 2L (liquid) where the Li layer is potentially disordered and not all Li sites are occupied.
As a result, the Li content $x$ can be between $0.33$ (stage 2L) and $0.50$ (stage 2) for a similar staging order.\cite{missyul_xrd_2017} 
Ideally, one would be able to fully resolve all the crystalline structures present in the sample at any given time and thus quantify the Li content, using precise Rietveld refinements with either X-rays\cite{missyul_xrd_2017} or neutron diffraction.\cite{taminato_real-time_2016,didier_phase_2020}
However, this approach requires excellent signal-to-noise ratio (SNR) over an extended $q$-range, which is hardly compatible with fast \textit{operando} experiment and may still be limited by the intrinsic disorder of some of the 
phases.
Another alternative approach is to use a stroboscopic technique, \textit{i.e.} repeated short illuminations with a controlled delay over several cycles to increase the SNR, as successfully performed by Sheptyakov \textit{et al.}  at the electrode scale using neutron diffraction.\cite{sheptyakov_stroboscopic_2020} 
It is yet to be verified that the underlying hypothesis of reproducible cycling still holds at the local scale.
Conversely, one could simply consider only the Bragg reflection from each stage and somewhat arbitrarily attribute it to a given composition, with the limitations that have been stated before.\cite{yao_operando_2019,berhaut_multiscale_2019,berhaut_prelithiation_2020}
Here we adopted an alternative strategy based on an empirical law and described hereafter.
We considered only the diffracted signal between the LiC$_{6}$ 001 Bragg reflection and the graphite 002 ($1.70 \leq q \leq 1.87~\text{\AA}$), \textit{i.e.} only the reflection on adjacent graphene sheets.
We suggest that we can estimate the local average $x$ composition from a  relation between the intensity-weighted average $q$ value over the whole $q$ range (or conversely d-spacings using $d = 2 \pi / q$) and the local average Li concentration $x$ in Li$_{x}$C$_{6}$, as described hereafter.
In figure \ref{fig:Ech1_deLi1_x_function_of_d} we have plotted the compositions $x$ as a function of the $q$-value from several reports in the literature over the past 45 years.
All the reported data are quite consistent, some uncertainty remains on the exact $q$ value for LiC$_{6}$ but most importantly, in the region $0.25 \leq x \leq 0.5$ (\textit{i.e.} Li$_{x}$C$_{24}$ to Li$_{x}$C$_{12}$), there is a very limited dependence of the $q$ value on the composition $x$, \textit{e.g.} the stage 2 and 2L mentioned above.
Nonetheless, we attempted to fit a piecewise linear function $f$ to establish the aforementioned relation between $q$ and $x$ and obtained the result shown in figure \ref{fig:Ech1_deLi1_x_function_of_d}.
The details of the fit are given in the Supplementary Information

For each $z$ location that we have measured, we calculated the intensity-weighted average $q$ value over the whole $q$ range between LiC$_{6}$ 001 and graphite 002. 
Prior to the intensity weighting, the intensity was corrected by the structure factor to take into account the reduction of the scattered amplitude by the inserted Li layer, which is scattering in antiphase with the graphene sheets.
Volume changes for the different Li$_\text{x}$C$_{6}$ compositions, \textit{i.e.} the shrinking of the electrode upon delithiation were also considered.\cite{missyul_xrd_2017}
The corrected-intensity-weighted average $q$-value at each location was then used to calculate the local average $x$ value  using the $f$ function introduced above.
The aforementioned limitations must be kept in mind, it is clear from the data scattering and the almost vertical slope around $0.25 \leq x \leq 0.50$ that the quantification of the Li content in this region is difficult.
In order to estimate the robustness of our approach, we tested different fitting functions $f$ and considered all those with a geometric distance to the data points smaller than twice the best fitting function (see SI).
We could thus test the dependency of our results on the fitting function and estimate the uncertainty of our approach.

\begin{figure}
    \centering
    \includegraphics[scale=1.0]{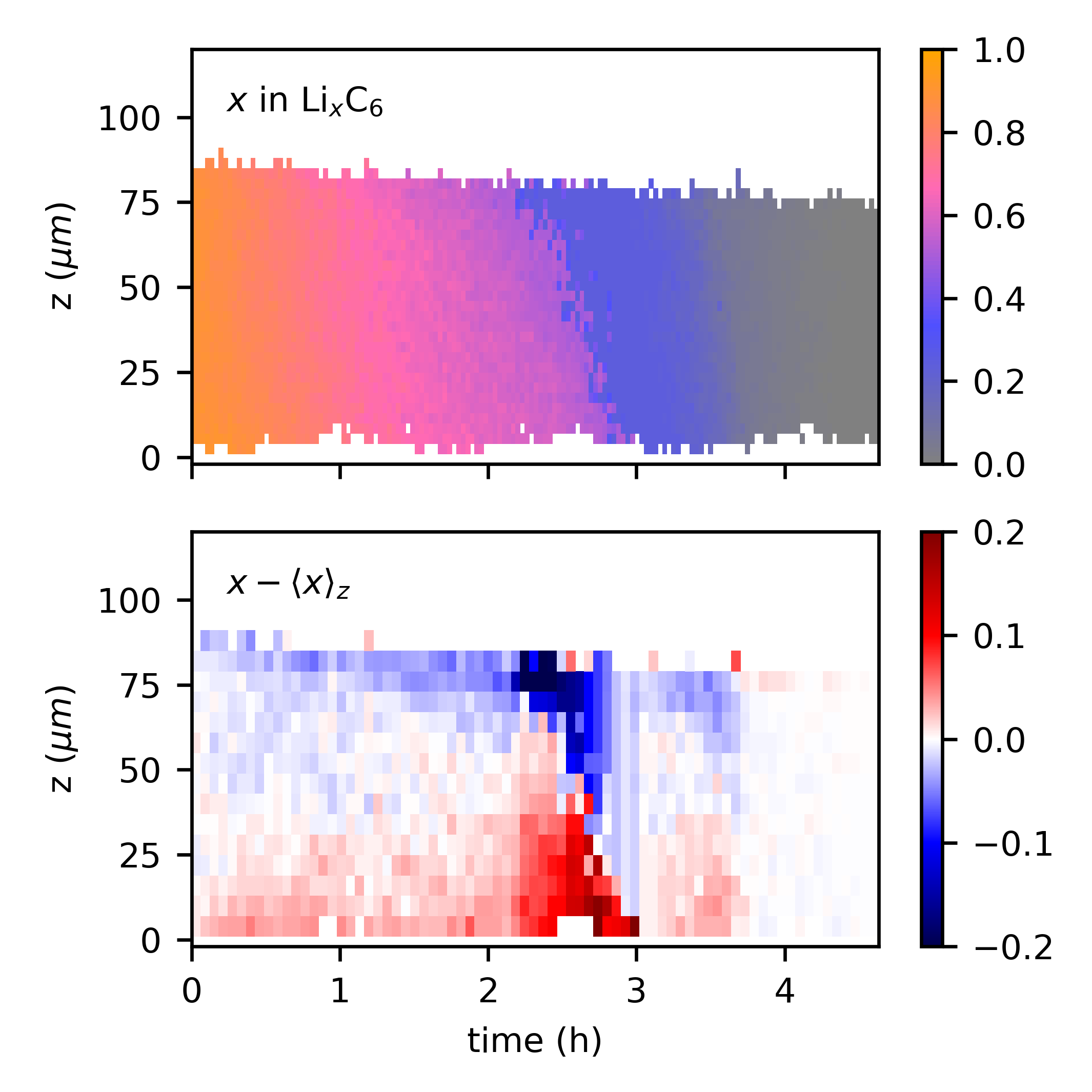}
    \caption{\label{fig:Ech1_deLi1_x_and_x-xmean_rebin}(top) Computed distribution of the Li content $x$ across the thickness of the electrode during the delithitiation, (bottom) deviation of the Li content to the mean value along z for each scan, indicated by its acquisition time. Note that the bottom panel has undergone a $2\times2$ rebin to reduce the noise.}
\end{figure}

The resulting map of the electrode composition $x$ across the electrode is plotted in figure \ref{fig:Ech1_deLi1_x_and_x-xmean_rebin}.
We can observe the depth-resolved evolution of the Li content as a function of time. 
We observe a smooth transition from $x\approx1$ to $x=0$ over the whole electrode, with a sharper transition around $x\approx0.4$ (typically around the stage  $2 \rightarrow 3$ transition), due to the steep slope of the $f$ function around these values.
Though it is clear that the exact $x$ value at that transition is highly dependent on the choice of $f$, there is a noticeable difference in the average $x$ value between the top and the bottom of the electrode, indicating a heterogeneous distribution of the composition.
Heterogeneities in electrode lithiation across the thickness are better seen in the bottom panel, were the difference to the mean value along $z$ is reported.
It highlights the heterogeneous stage  $2 \rightarrow 3$ transition, and stage $3 \rightarrow 1_{\mathrm{d}} + \mathrm{G}$ transition, at around $2.6$~h and $3.5$~h, respectively. %scan $\#$85 and $\#$115 respectively.
Around such times, the top of the electrode, near the separator, is systematically more delithiated than the bottom, near the current collector.

Our estimate of the Li content $x$ can also be double-checked by comparing it to the estimate from the macroscopic capacity measurement.
We obtain a fair agreement over the whole delithiation process, as shown in figure S3 of the Supplementary Information %\ref{fig:Ech1_deLi1_x_compare_ec_xrd}. 
Slight discrepancies can be seen for $x$ between 0.25 and 0.55, with an overestimation followed by an underestimation.
This most likely stems from the difficult separation of the phases with those compositions with X-ray diffraction.
Differences observed at low $x$ values may be due to parasitic reactions or the lithiation of disordered graphite, which is not visible in diffraction.
We also note that our quantification and its comparison to the capacity measurement is quite similar to that obtained by Yao \textit{et al.} using a decomposition of the same diffraction peaks over a finite set of five stages.\cite{yao_operando_2019}

\begin{figure}
    \centering
    \includegraphics[scale=1.0]{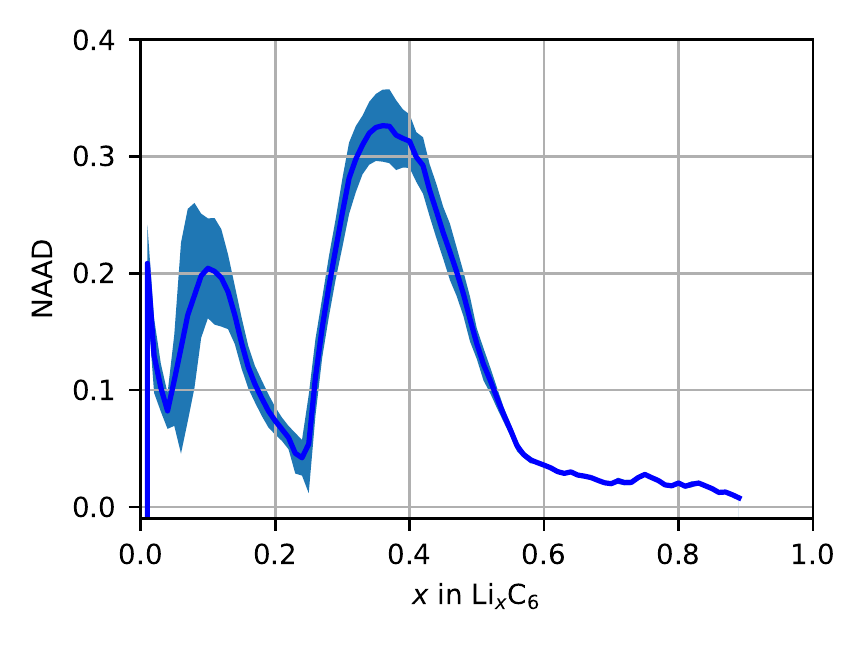}
    \caption{\label{fig:Ech1_deLi1_NAAD}NAAD of the Li content $x$ as a function of $x$, from the analysis of the diffraction experimental results. 
The solid line corresponds to the calculation using the best fit from figure \ref{fig:Ech1_deLi1_x_function_of_d}, while the shadowed area is the estimated error.
During the experiment, $x$ goes from $\approx 0.9$ to $0$, \textit{i.e.} from an almost fully lithiated to fully delithiated electrode.}
\end{figure}

We can now focus on the quantification of heterogeneities and compute the experimental NAAD from the quantified Li content $x$, as introduced in the model.
Figure \ref{fig:Ech1_deLi1_NAAD} shows the experimental NAAD and the estimated error.
We recover some of the striking features predicted by the simulation, including two large bumps around $x=0.10$ and $x=0.37$, two minima around $x=0.04$ and $x=0.22$, and a slower decrease for $x\geq0.55$. 
The positions of the extrema are quite robust and insensitive to the choice of the $f$ function used to treat the data, while the amplitude of the maxima can show some variations. 
These extrema positions are very consistent with the model predictions, and well correlated with the graphite equilibrium potential shape. 
Similar peculiar NAAD shape at stoichiometry below $x \leq 0.5$ was also observed for two other conditions (see Supplementary Information), hence demonstrating the reliability of the data acquisition, treatment and analysis.
 
Contrary to the simulations, no significant NAAD increase is however observed at higher stoichiometry, \textit{i.e.} around the stage $1 \rightarrow 2$ transition. 
As seen in figure \ref{fig:Ech1_deLi1_phases_maps} and previously described, this transition seems really homogeneous throughout the electrode thickness. 
Thus, while the stage $1 \rightarrow 2$ two-phase transition is associated with a plateau on the graphite equilibrium potential and should promote heterogeneous lithiation along the thickness, no such behavior is experimentally observed.
Therefore it implies revising the description of the limiting phenomena during this particular transition in our model. 

\section{Discussion}
To analyze the discrepancy observed at larger $x$ values, it is necessary to revisit the model. 
Newman-type models with solid-solution approach for lithium transport inside the active particles do not allow to capture phase front propagation in multi-phase materials such as graphite, but remain sufficient to highlight the main competing phenomena and their relative weights. 
The competition between lithium transport through the electrode thickness and lithium intercalation inside graphite drives lithium heterogeneities at the electrode scale. It is clear from the present experimental results that Newman-type models can predict accurately the succession of homogeneous/heterogeneous distribution of lithium content across the electrode thickness for the lower half of the $x$ regime.    

\begin{figure}
    \centering
    \includegraphics[scale=1]{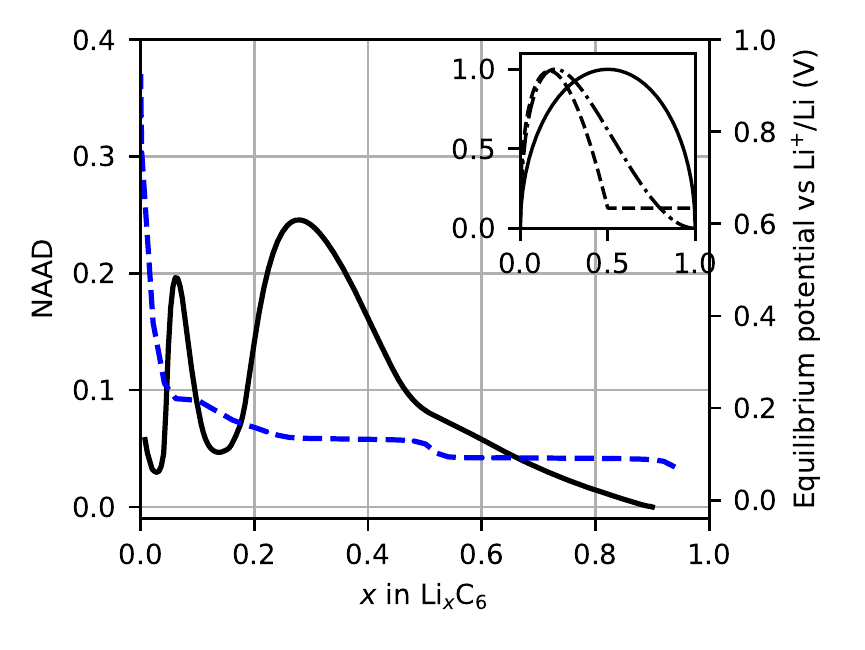}
    \caption{\label{fig:NAAD_Simu} Simulated NAAD of the volume Li content $x_{Li}$ inside the graphite particles as function of the stoichiometry $x$ in Li$_x$C$_6$ obtained with the modified Butler-Volmer model (solid line) and corresponding simulated equilibrium potential of graphite (dashed lines). (insert) Dimensionless exchange current densities for the traditional Butler-Volmer kinetics (solid line), Bazant theory for LFP (dashed-dotted line), and proposed modified expression (dashed line)}
\end{figure}

The main driving processes, \textit{e.g.} Li transport through the electrode and Li intercalation in graphite, seem to be differently balanced at high stoichiometry.
%The balance between lithium transport through the electrode and intercalation seems different at high stoichiometry. 
The experimentally observed homogeneous distribution of lithium during the stage $1 \rightarrow 2$ transition indicates that lithium transport through the electrode thickness is not the limiting phenomena compared to lithium (de)-intercalation during this transition. 
Since lithium transport in the electrolyte is not impacted by the stoichiometry, we must reconsider the lithium intercalation description. 
The kinetics of lithium intercalation is driven by the exchange current density (ECD) at the surface of the graphite particle and by the lithium transport inside them.
Since the diffusion approach used in the present porous electrode model is anyway not suited to describe transport in materials undergoing phase transitions, we do not modify it and keep it as simple as possible.
Therefore, we focus on the description of the ECD.
In our initial work following classical descriptions,\cite{Doyle1993} we used a Butler-Volmer expression to model the ECD, and took it proportional to $\left(1-x^s_{Li}\right)^{(1-\alpha_{gr})} {x^s_{Li}}^{\alpha_{gr}}$, where $\alpha_{gr}$ is the so-called asymmetry coefficient and $x^s_{Li}$ is the lithium stoichiometry at the surface of the graphite particle. 
A value of $\alpha_{gr}=0.5$ leads to a symmetric ECD, with a maximum at $x=0.5$ and two equal minimum at $x=0$ and $x=1$.
As underlined by Bazant,\cite{Bazant2013} the classical Butler-Volmer model does not account for the activity coefficient of the transition state of the reaction. 
Introducing a modified expression of the ECD to describe the exclusion of intercalation sites during the reaction, the position of the maximum can be shifted to lower stoichiometry.
This leads to an auto-inhibitory or auto-catalytic lithium intercalation or deintercalation, as discussed by Bazant.\cite{bazant_thermodynamic_2017}
In particular, this shift of the maximum of the ECD reduces the kinetics of (de)intercalation at high stoichiometry.
This is consistent with our experimental results, since lower (de)intercalation kinetics leads to lower heterogeneities. 
Nevertheless, we could not reproduce the striking experimental features of lithium distribution inside the graphite electrode using the modified expression introduced by Bazant \textit{et al.}~\cite{Bazant2013,Cogswell2012,Bai2011} to study phase separation in LiFePO$_4$ (see insert in figure~\ref{fig:NAAD_Simu}). 
We found that a much lower value of the (de)intercalation kinetics is necessary for $x \geq 0.5$ to wash out the heterogeneities only at high stoichiometry. 
Considering such a form of the ECD, with a much lower value for $x\geq0.5$ (insert in figure~\ref{fig:NAAD_Simu}), we could obtain a calculated NAAD in agreement with the experimental observations, namely the two low-$x$ heterogeneities are maintained, while the one at large $x$ disappears, as can be seen figure~\ref{fig:NAAD_Simu}. 

Further theoretical work would be necessary to propose an ECD expression with more sound thermodynamic arguments and a clear relation with phase front propagation, which is beyond the scope of this work. 
We can nevertheless highlight that an ECD with a much lower value during the LiC$_6$/LiC$_{12}$ phase transition is required to be consistent with the experimental observations. 
This drastic reduction of the ECD could be related to the higher repulsive energies between the intercalated lithiums at high stoichiometry, which impacts the intercalation kinetics. 

Low values of the exchange current density, not only close to $x=1$ as in the Butler-Volmer expression, but for all $x \geq 0.5$, could imply revisiting our understanding of lithium plating, a phenomenon that occurs during lithiation of the negative electrode when the rate of lithium deposition exceeds the rate of lithium intercalation.  
Yao \textit{et al.}\cite{yao_quantifying_2019} already highlighted the importance of accounting for lithium concentration gradients through the electrode and not just averaged values to correctly capture the occurrence of this degradation phenomena. 
The low values of the ECD during all the stage $1 \leftrightarrow 2$ transition reported in this work implies important over-potentials inside the graphite electrode even at low C-rates and not only close to the end of lithiation, but during a much longer period. 
This finding may have important consequences as negative graphite electrode potentials, which favor the formation of lithium deposit, would be reached for much lower electrode thickness and C-rate than what is currently predicted with models using  classical Butler-Volmer expression for the ECD.\cite{Hein2016,Yang2017}
Further investigations on the intercalation kinetics of lithium in graphite as a function of the stoichiometry seem therefore crucial for better optimization of graphite electrode under fast charge conditions.      

\section{Conclusions}
Graphite remains the most widely used active material within lithium-ion negative electrodes.
Detailed understanding of lithium dynamics inside negative electrodes is critical in optimizing these electrodes in terms of performance, lifetime and safety. 
In this study, lithium distribution across a $80~\mu m$ thick porous graphite electrode has been followed \textit{operando} using  $1~\mu m$-spatially resolved X-ray diffraction during delithiation at a moderate C-rate (C/5).
Quantitative analysis of the lithium heterogeneities that develop across the electrode thickness reveals a very particular pattern with alternating sequences of homogeneous and heterogeneous lithium distributions.  
For stoichiometries $x \leq 0.5$, the main striking features of the experimental pattern match the prediction from our previous numerical study based on a Newman-type porous electrode model. 
Thus, the experimental approach developed here validates for the first time such modeling studies at the local scale, \textit{i.e.} across the electrode thickness. 
It also strengthen our analysis of lithium heterogeneities based on the competition between lithium transport through the electrode and variation of graphite equilibrium potential with stoichiometry. 

At higher stoichiometries, \textit{e.g.} during the stage $1 \rightarrow 2$ transition, no significant heterogeneities are experimentally observed, contrary to the model predictions.
This observation implies that lithium transport through the electrode thickness is not the limiting phenomena during this two-phase transition, and assumptions of the model must be refined.
Accordingly, we revisited the model and show that a non-symmetric exchange current densities with a very low value for $x \geq 0.5$ can explain the homogeneous behavior observed during the stage $1 \rightarrow 2$ transition. 
Further theoretical work is needed to understand the grounds of such a non-symmetric behavior. 

The present experimental methodology allows the quantification of lithium intercalation with both space and time resolution. 
It is therefore a high-value tool to validate and improve the theoretical models of the electrodes, which are needed for materials and cell developments.   
Furthermore, the combination of modeling and advanced \textit{operando} synchrotron experiment that we present here is extendable to other anode and cathode intercalation materials, including next generation energy storage technologies, beyond LiB. 

\section*{Conflicts of interest}
There are no conflicts to declare.

\section*{Acknowledgements}
We are grateful to Peter van der Linden and Diego Pontoni from the PSCM Group at the ESRF for 3D printing parts for the experimental setup, and Philippe Montmayeul at CEA for machining the electrochemical cell. 
We acknowledge the use of the EC-LAB facility at the ESRF, and the Hybrid'En and LITEN batteries platform at CEA to prepare the cells prior to the experiments. 

\bibliography{article_arXiv_ID13_Tardif} 
\bibliographystyle{unsrt}

\setboolean{@twoside}{false}


\section*{Supplementary material}

# Contribution of operando X-ray experiments and modeling to the understanding of the heterogeneous lithiation of a graphite electrode

## SUPPLEMENTARY INFORMATION

Samuel Tardif,*[a] Nicolas Dufour,[b] Jean-Francois Colin,[b] Gérard Gébel,[b] Manfred Burghammer,[c] Andreas Johannes,[c] Sandrine Lyonnard,[a] and Marion Chandesris[b]

[a]Department of Physics, IRIG, Univ. Grenoble Alpes and CEA, F-38000 Grenoble, France ; E-mail: samuel.tardif@cea.fr
[b]Department of Electricity and Hydrogen for Transport, Univ. Grenoble Alpes and CEA-LITEN, F-38000 Grenoble, France
[c]European Synchrotron Radiation Facility (ESRF), BP 220, Grenoble 38043, France

## Contents

### 1. Formation of the electrode (first cycle)

The electrode was prepared and cycled just before the beamtime, as shown in Figure S1. The sequence applied was:

1. lithiation for 10 h (constant current of 60 µA)
2. floating for 1 h at 5 mV
3. relaxation for about 15 min (open circuit)
4. delithiation until the cutoff potential of 1.5 V (constant current of 60 µA for 9h17m)
5. floating for 5 min (constant potential of 1.5 V)
6. relaxation for about 15 min
7. second lithiation for about 8h (constant current of 75 µA)
8. floating for 2h45 at 5 mV
9. relaxation for 30 min
10. open circuit during transfer to the beamline

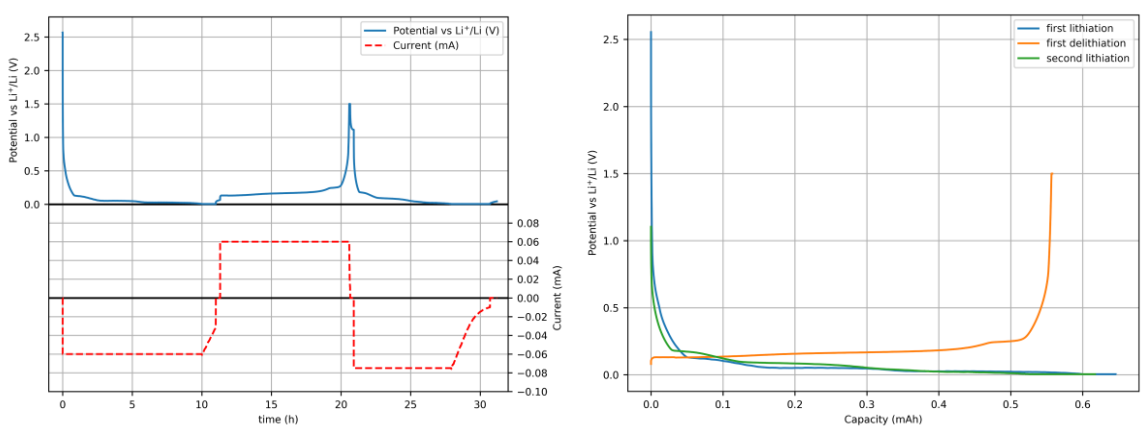

*Figure S1: Potential of the operando during the preliminary cycling of the electrode, prior to the experiment, (left) as a function of time, (right) as a function of capacity.*

## 2. Quantification of the Li content from X-ray diffraction

As described in the text, we tested different fit function to establish a relation between the average Li content $x$ and the average $q$ value, or correspondingly $d$-spacing ($d = \frac{2\pi}{q}$), of the diffraction from adjacent graphene sheets. We first found the best fit for a piecewise linear function going through the [q,x] points listed below, ordered by increasing $x$ (and $d$) :

[$q_0$ = 1.873 Å$^{-1}$, $x_0$ = 0], [$q_1$, $x_1$], [$q_2$, $x_2$], [$q_3$, $x_3$ = 0.25], [$q_4$, $x_4$ = 0.5], [$q_5$ = 1.701 Å$^{-1}$, $x_5$ = 1],

solving for $q_1$, $q_2$, $q_3$, $q_4$, $x_1$, and $x_2$, with the condition that $q_1 > q_2 > q_3 > q_4$ and $x_1 < x_2$.

The extreme $q$, $x$ values ($q_0$, $x_0$ and $q_5$, $x_5$) are given by our experiment: the peak from LiC$_6$ ($x_5$ = 1) is well separated from the other possible contributions and the $q$ value is unambiguous, and the $q$ value for the peak from graphite ($x_0$ = 0) is the asymptotic position at the end of delithiation. This value may possibly be affected by some graphite that would somehow have remained partially lithiated but from the intensity map in the main text, the error is clearly limited. Note that both $q_0$ and $q_5$ values fall well within reported values for the same compounds. We fixed $x_4$ = 0.5, where a strong change of slope can be observed in the data from the literature. No clear change of slope could be observed around $x$ = 0.33 (LiC$_{18}$), so we fixed the next value $x_3$ to 0.25 (LiC$_{24}$).

The positions of the points for the best *f* fit function are reported in Table S1 (fixed values are in italic). From these values, we tested other *f\** functions at +/- 0.004 Å (in *d*-spacing) and +/- 0.02 (in *x* value). The geometric distance between these new *f\** functions and the data from the literature was computed. All *f\** within twice the minimum distance were kept (Figure S2), and *in fine* used to estimate the NAAD uncertainty in Figure 9.

2

| Point | *d* (Å) | *q* (Å⁻¹) | *x* |
|---|---|---|---|
| 0 | *3.354* | *1.873* | *0.000* |
| 1 | 3.431 | 1.831 | 0.066 |
| 2 | 3.491 | 1.800 | 0.250 |
| 3 | 3.518 | 1.786 | 0.250 |
| 4 | 3.521 | 1.785 | *0.500* |
| 5 | *3.693* | *1.701* | *1.000* |

*Table S1: Position of the (q,x) points defining the piecewise function f used in the text. Values in italic were held for the fit.*

The structure factor correction was as follows. For a given *f\** function, we calculated for each *q*-value the corresponding *x* value using *f\**. Then we estimated the structure factor F considering a 180° phase shift between the graphene sheets and the Li plane, using the atomic scattering factors from the International Tables of Crystallography, interpolated over the *q* range of interest. The volume correction was simply estimated supposing a linear increase of the volume V with *x*. The experimental intensity was then divided by $|F^2|/V^2$.

Using the corrected intensity, we calculated the intensity-weighted average *q* value at each measurement location, and thus estimate the local average *x* using *f\**. The profile of *x(z)* could then be obtained and used to calculate the NAAD using equation 1 from the main text.

All the NAAD profiles using the different *f\** functions were calculated, and the mean NAAD is plotted in Figure 9 in the main text. The shaded area corresponds to +/- 1 standard deviation across all the *f\** functions used and provides an estimate of the uncertainty.

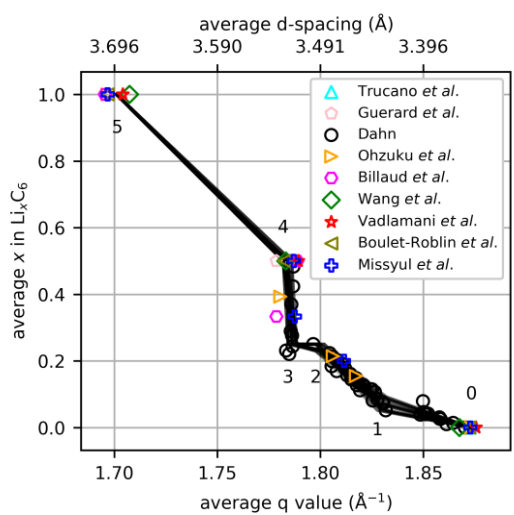

*Figure S2: Different x=f(q) relations tested to estimate the robustness of the NAAD calculation and corresponding errors. Symbols are taken from the literature. References are available in the article. The solid line is least-square fit with a piecewise linear function.*

### 3. Comparaison of the Lithium content: XRD *vs* Electrochemistry

In order to assess the quality of the XRD quantification discussed above, we compared the mean value of *x* over the whole electrode (*i.e.* the state of charge) as calculated from the XRD, and as measured with the galvanostat. The results are shown in figure S3. Differences are within ±0.1 and maybe explained by the fact that XRD is only sensitive to the ordered intercalation compounds, and that the electrochemical capacity measurements can also include electric current from other side reactions, such as SEI formation.

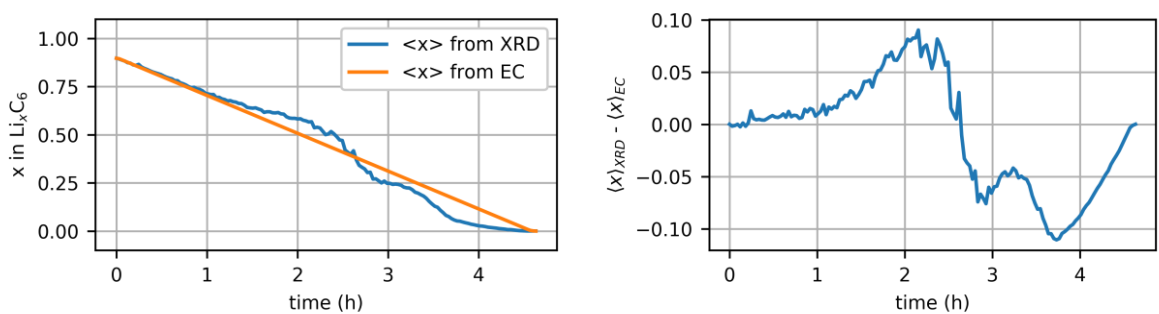

*Figure S3: (left) Comparison of the Li content estimated from the XRD and electrochemical capacity (EC) measurements; (right) difference between the lithium content x measured using the microscopic XRD and the macroscopic electrochemistry as a function of the x (from the XRD).*

### 4. Experimental reproducibility

We were able to test the reproducibility of our observation in two other instances. Firstly we lithiated again the same sample (Sample1_Li2 in Figure S4), and secondly we performed the initial lithiation in another similar sample (Sample2_Li1 in Figure S4). In both cases we observed the particular succession of homogeneous and heterogeneous distribution of Li content across the electrode depth, as shown by the local increases in NAAD in figure S4. Note that contrary to the case exposed in the main text (*i.e.* delithiation, *x* starting near 1 and decreasing with time), these additional measurements were performed in lithiation (*x* starting near 0 and increasing with time). Due to time constraints, in both cases we could not reach the higher stoichiometries.

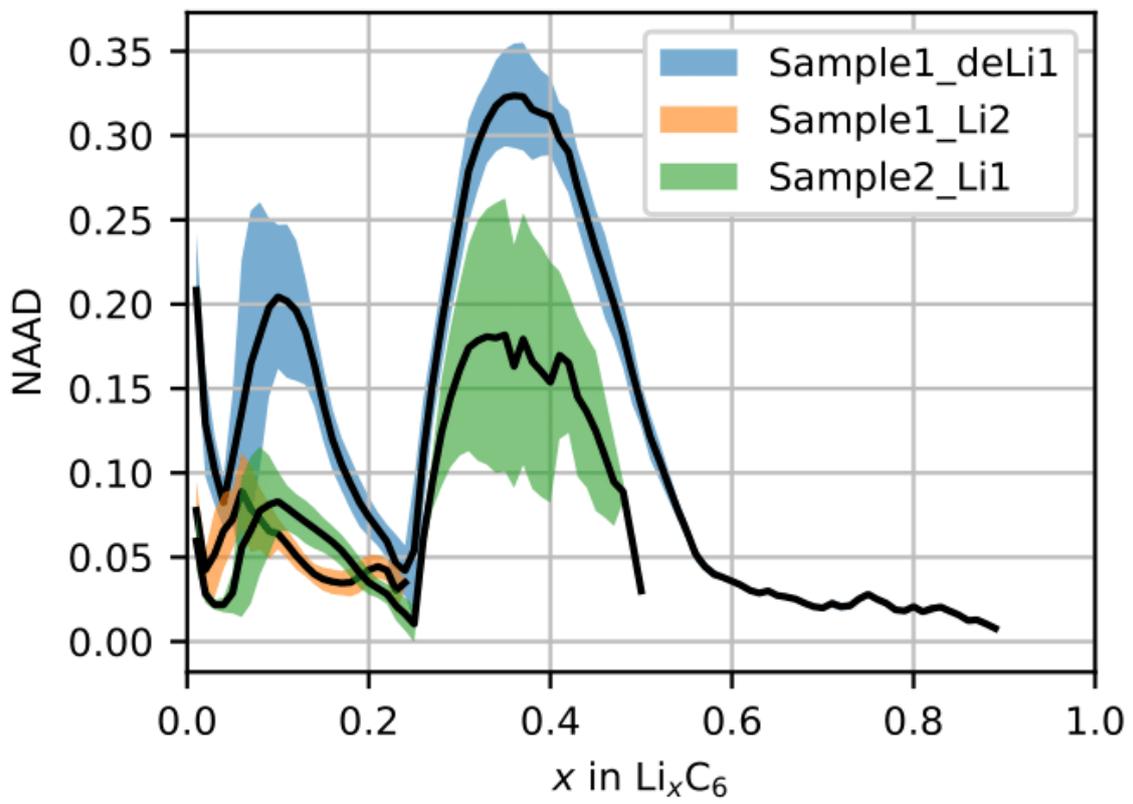

*Figure S4: Comparison of the NAAD as function of the lithium stoichiometry in LixC6 for different conditions: delithiation of sample 1, following delithiation of the same sample and first lithiation of a second sample.*

## 5. Porous Electrode Model

The graphite electrode model is based on the porous electrode theory [1]. The resolved equations are classical and a summary is provided in table S2. A more detailed description of the hypothesis is available in one of our previous work [2]. All the simulations were performed at a C-rate of C/5.

| Graphite electrode and separator domains - Liquid phase (electrolyte) | | |
|---|---|---|
| Mass balance | $\epsilon \dfrac{\partial C_l}{\partial t} - \nabla \cdot (D_l^{eff} \nabla C_l) = \dfrac{j_l}{F}(1 - t_0^+)$ | |
| Charge balance | $\nabla \cdot \left( -\sigma_l^{eff} \left( \nabla \phi_l - \dfrac{2RT}{F}(1 - t_0^+)(1 + \dfrac{dlnf^{\mp}}{dlnc}) \dfrac{\nabla C_l}{C_l} \right) \right) = j_l$ | |
| Currents for electrode and separator | In the graphite electrode $j_l = -j_s$ | In the separator $j_l = 0$ |
| **Graphite electrode domain - Solid phase.** | | |
| Charge balance | $\nabla \cdot (-\sigma_s^{eff} \nabla \phi_s) = j_s$ | |
| Currents for electrode | $j_s = a_{Gr} i_{Gr}$ | |
| **current at the active materials and electrolyte interface (Notation: $i_n = -i_{Gr}$)** | | |
| Current density at the active material/electrolyte interface | $i_n = i_o \left( exp\left(\dfrac{\alpha \eta F}{RT}\right) - exp\left(-\dfrac{(1-\alpha)\eta F}{RT}\right) \right)$ | |
| Overpotential | $\eta = \phi_s - \phi_l - E_{eq}(x_{Li}^s, T)$ | |
| Nominal current density | $i_o = i_{o,ref} (1 - x_{Li}^s)^{(1-\alpha)} (x_{Li}^s)^{(\alpha)} \left(\dfrac{C_l}{C_{l,ref}}\right)^{\alpha}$ | |
| **active particles domains** | | |
| Mass balance | $\dfrac{\partial C_s}{\partial t} - \nabla \cdot (D_s \nabla C_s) = 0$ | |

*Table S2: Governing equations of the graphite porous electrode model.*

The effective properties in the porous graphite electrode and separator are computed from the porosity $\epsilon$ and tortuosity $\tau$ values for each domain: $D_l^{eff} = D_l \epsilon / \tau$ and $\sigma_l^{eff} = \sigma_l \epsilon / \tau$.

The physical parameters used in the simulations are presented in the tables below, components by components.

| Parameter | Symbol | Value | Note |
|---|---|---|---|
| Derivation of the activity coefficient | $\dfrac{dlnf^{\mp}}{dlnc}$ | 0 | Hypothesis |
| Transference number | $t_0^+$ | 0,363 | From ref [3] |
| Concentration LiPF$_6$ | $C_{l,ref}$ | 1 mol/L | From ref [2] |
| Lithium diffusivity | $D_l$ | $5.10^{-11}$ m²/s | From ref [2] |
| Ionic conductivity | $\sigma_l(C_l)$ | $1.73 + 17.92 C_l + 13.98 C_l^2 + 2.67\, C_l^{\,3}$ | From ref [4] |

Table S3: Electrolyte properties used in the model

| Parameter | Symbol | Value | Note |
|---|---|---|---|
| Separator porosity | $\epsilon$ | 0.41 | From ref [2] |
| Separator tortuosity | $\tau$ | 2.67 | From ref [2] |
| Separator thickness | L | 50µm | From ref [2] |

Table S4: Separator properties used in the model

| Parameter | Symbol | Value | Note |
|---|---|---|---|
| Composition | | 96 %wt Graphite / 4%wt Binder; | |
| Graphite weight fraction | $W_{Gr}$ | 0.96 | From composition |
| electrode porosity | $\epsilon$ | 0.35 | From ref [2] |
| electrode tortuosity | $\tau$ | 4 | From ref [2] |
| electrode thickness | L | 84.2µm | From ref [2] |
| electrode effective conductivity | $\sigma_s^{eff}$ | 100S/m | From ref [4] and [2] |
| Electrode density | $\rho$ | 2236kg/m³ | From ref [2] |

Table S5: Graphite electrode properties used in the model

| Parameter | Symbol | Value | Note |
|---|---|---|---|
| Mean particle radius | $R_{Gr}$ | 8μm | From ref [2] |
| Graphite density | $\rho_{Gr}$ | 2260 kg/m$^3$ | From ref [2] |
| Reversible capacity | $q_{Gr}$ | 355 mAh/g | From ref [2] |
| Maximum lithium concentration | $C_{s,max}^{Gr}$ | 31 370 mol/m$^3$ | computed |
| Equilibrium potential | $E_e$ | See figure 1 in main text | From ref [2] |
| Diffusion coefficient of lithium in graphite | $D_s$ | 5.10$^{-13}$ m²/s | From ref [4] |
| Symmetry coefficient for BV relation | $\alpha$ | 0.5 | Hypothesis, also ref[2] |
| Nominal exchange current density | $i_{o,ref}^{Gr}$ | 4.7 A/m² | Adjusted |

*Table S6: Graphite properties used in the model*



\end{document}